\documentclass[twocolumn]{article}
\usepackage{fullpage}
\usepackage[T1]{fontenc}
\usepackage[utf8]{inputenc}
\usepackage{lmodern}
\usepackage{textcomp}
\usepackage{lastpage}
\usepackage{booktabs}
\usepackage{comment}
\usepackage{hyperref}
\usepackage{graphicx}
\graphicspath{ {./images/} }
\usepackage{listings}
\usepackage{longtable}
\usepackage{listings}
\usepackage{caption}
\usepackage{subcaption}
\usepackage{float}
\usepackage{titlesec}

\titleformat{\paragraph}
{\normalfont\normalsize\bfseries}{\theparagraph}{1em}{}
\titlespacing*{\paragraph}
{0pt}{3.25ex plus 1ex minus .2ex}{1.5ex plus .2ex}

\newcommand{\TITLE}{ 
Approaches to the Parallelization of \\
Merge Sort in Python}

\newcommand{\AUTHOR}{Alexandra Yang}

\title{\TITLE}
\author{\AUTHOR}
\date{\centering drayales@gmail.com}

\begin{document}

\onecolumn
\begin{center}
{\huge \TITLE}

{\AUTHOR}
\end{center}

\tableofcontents

\twocolumn

\newpage

\maketitle

\begin{abstract}

The theory of divide-and-conquer parallelization has been well-studied \cite{selkie} in the past, providing a solid basis upon which to explore different approaches to the parallelization of merge sort in Python. Python's simplicity and  extensive selection of libraries make it the most popular scientific programming language, so it is a fitting language in which to implement and analyze these algorithms.

In this paper, we use Python packages multiprocessing and mpi4py to implement several different parallel merge sort algorithms. Experiments are conducted on an academic supercomputer, upon which benchmarks are performed using Cloudmesh. We find that hybrid multiprocessing merge sort outperforms several other algorithms, achieving a 1.5x speedup compared to the built-in Python sorted() and a 34x speedup compared to sequential merge sort. Our results provide insight into different approaches to implementing parallel merge sort in Python and contribute to the understanding of general divide-and-conquer parallelization in Python on both shared and distributed memory systems. 

\end{abstract}

\section{Introduction}

Divide-and-conquer algorithms recursively break problems down into smaller subproblems until those become simple enough to solve directly, at which point the recursion terminates. The results are then combined to generate a solution to the original problem. Divide-and-conquer algorithms are widely used for many different purposes, including multiplying very large numbers, finding the closest pair of points, and computing the discrete Fourier transform. 

Divide-and-conquer algorithms are uniquely suited to parallelization. Parallelism is usually inherent, as once the division phase is complete, the subproblems are usually independent from one another and can be solved separately. Furthermore, phases often mirror each other, with the combination phase of one subproblem occurring identically, but independently, to the combination phase of another subproblem. 

Merge sort is the classic divide-and-conquer algorithm, making it an excellent experimental base upon which to try different approaches to parallelism. In this paper, we explore the parallelization of merge sort using several different Python packages. To parallelize merge sort on Symmetric Multi-Processors (SMPs), we use multiprocessing, and to parallelize it on clustered systems, we use message-passing through the mpi4py module. 

We begin by reviewing several single-node sorting algorithms in Section \ref{single-sorting}. Section \ref{multiprocessing-ms} explores a shared memory multiprocessing implementation of merge sort. We evaluate and analyze its performance in comparison with sequential merge sort (Section \ref{seq-ms}) and the built-in Python sort (Section \ref{builtin}) in Section \ref{analysis1}.

Next, we describe two merge sorts that are implemented using MPI. Section \ref{mpi-ms} focuses on both message-passing merge sort using mpi4py and a hybrid MPI merge sort that combines both multiprocessing and MPI. These algorithms are evaluated and compared in Section \ref{analysis2}. Finally, we offer our conclusions in Section \ref{end}. 

\section{Related Research}

The theory of merge sort parallelization has been well-studied in the past, with many different existing designs and implementations. Cole \cite{cole88} presents a parallel implementation of the merge sort algorithm with $O(log n)$  time complexity on a CREW PRAM, a shared memory abstract machine which does not provide synchronization and communication, but provides any number of processors. Another design is presented by Jeon and Kim \cite{jeon03}, who explore a load-balanced merge sort that evenly distributes data to all processors in each stage. They achieve a speedup of 9.6 compared to a sequential merge sort on a Cray T3E with their algorithm. In another theoretical design, Marszalek, Wozniak, and Polap \cite{mars18} propose a Fully Flexible Parallel Merge Sort with a theoretical complexity of $O(n(\log_2(n)^2/k)$. Both Uyar \cite{uyar14} and Odeh et al. \cite{odeh} propose new and innovative ways of the design of merge sort and implement them in a parallel manner, with Uyar proposing a double merging technique and Odeh et al. presenting a method of partitioning two sorted arrays. In contrast, Davidson et al. \cite{davidson} design and implement a high-performance merge sort for highly parallelized systems, which is run and benchmarked on GPUS. 

On MPI, Randenski \cite{radenski11}  describes three parallel merge sorts implemented in C/C++: shared memory merge sort with OpenMP, message passing merge sort with MPI, and a hybrid merge sort that uses both OpenMP and MPI. They conclude that the shared memory merge sort runs faster than the message-passsing merge sort. The hybrid merge sort, while slower than the shared memory merge sort, is faster than message-passing merge sort. However, they also mention that these relations may not hold for very large arrays that significantly exceed RAM capacity. This paper provides valuable information on different approaches to parallelization.

\section{Installation}

The source code is located in GitHub \cite{cloudmesh-mpi}. To install and run it, along with the necessary packages, run the following commands:

\begin{verbatim}
$ python -m venv ~/ENV3
$ source ~/ENV3/bin/activate
$ mkdir cm
$ cd cm
$ pip install pip -U
$ pip install cloudmesh-installer
$ cloudmesh-installer get mpi
$ cd cloudmesh-mpi
$ pip install mpi4py
$ pip install multiprocessing
\end{verbatim}

\section{Benchmarking}

The code uses the StopWatch from the Cloudmesh library, which easily allows code to be augmented with start and stop timer methods, along with the use of a convenient summary report in the format of a CSV table. \cite{las-stopwatch}. 

\section{Single Processor Sorting Algorithms} \label{single-sorting}

\subsection{Sequential Merge Sort} \label{seq-ms}

\begin{figure}[htb]
\includegraphics[width=1.0\columnwidth]{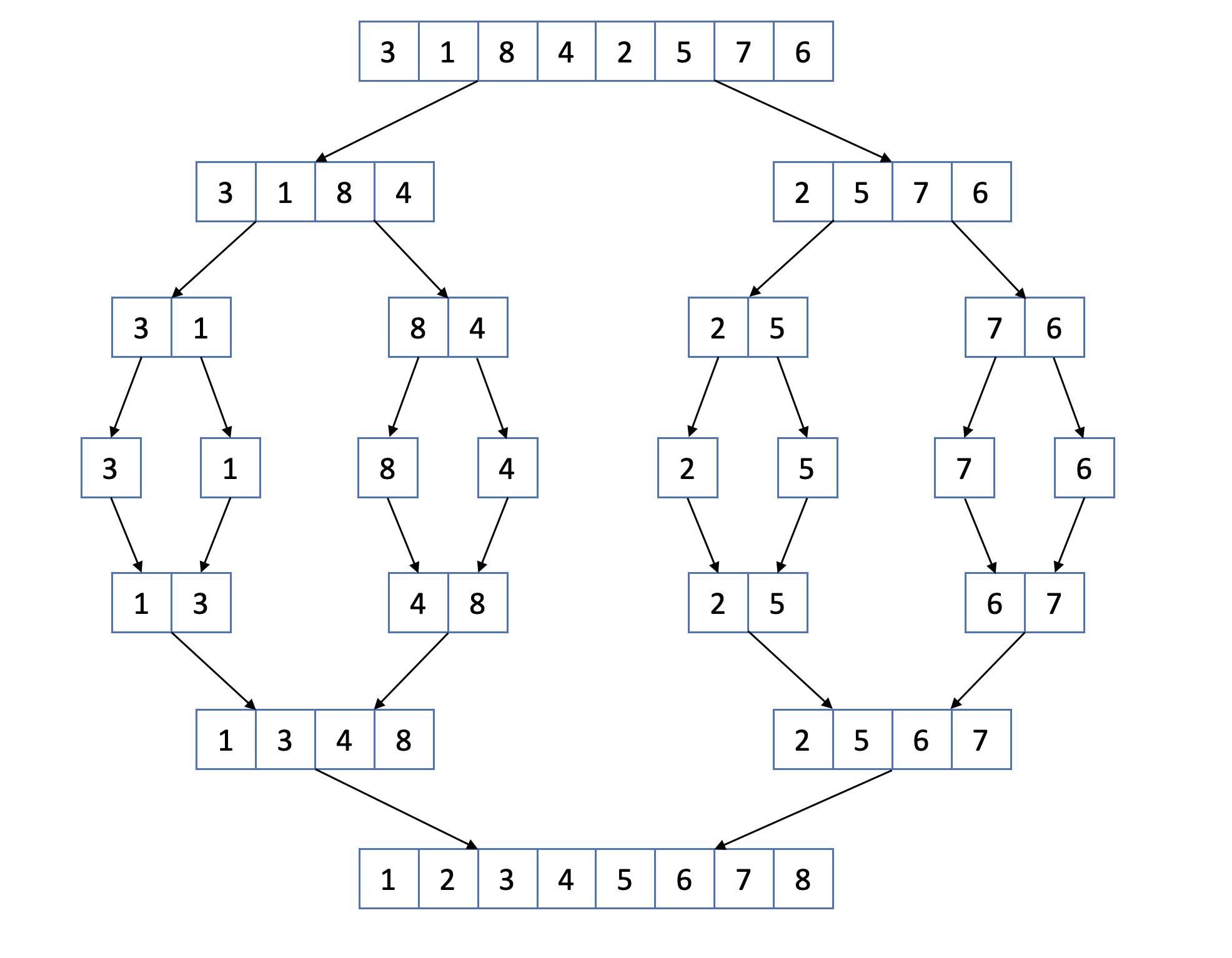}
\caption{Sequential merge sort example}
\label{fig:overview}
\end{figure}

Sequential merge sort is a standard example of a sequential divide-and-conquer algorithm. The idea of merge sort is to divide an unsorted list into smaller, sorted lists, and then merge them back together to produce the sorted list in its entirety (see Figure \ref{fig:overview}. 

The steps are as follows:

\begin{enumerate}
    \item If the input array has less than two elements, return.
    \item Split the input array into two halves, \lstinline{left} and \lstinline{right}. 
    \item Sort the two halves using the same algorithm. 
    \item Merge the two sorted halves to form the output array. 
\end{enumerate}

The code is presented in Figure \ref{fig:mergesort}

\begin{figure}[htb]
\footnotesize
\begin{lstlisting}
def mergesort(array):
  # check to make sure array has two or 
  # more elements
  if len(array) > 1:
    # find midpoint of array
    mid = len(array) // 2
            
    # split array into two halves
    left = array[:mid]
    right = array[:mid]
            
    # sort the two halves
    left = mergesort(left)
    right = mergesort(right)
            
    # merge the two halves
    array = merge(left, right)
  return array
\end{lstlisting}
\caption{Sequential merge sort algorithm.}
\label{fig:mergesort}
\end{figure}

A variant on this approach is to stop splitting the array once the size of the subarrays gets small enough. Once the subarrays get small enough, it may become more efficient to use other methods of sorting them, like the built-in Python \lstinline{sorted(a)} function instead of recursing all the way down to size 1. See Figure \ref{fig:mergesort-2}. 

\begin{figure}[htb]
\footnotesize
\begin{lstlisting}
def sequential_mergesort(array):
  n = len(arr)
    if n < SMALLEST_ARRAY_SIZE:
      array = sorted(array)
      return array
        
    else:
      left = array[:mid]
      right = array[:mid]
            
      # sort the two halves
      left = sequential_mergesort(left)
      right = sequential_mergesort(right)
            
      # merge the two halves
      array = merge(left, right)
        
      return array
\end{lstlisting}
\caption{Hybrid sequential merge sort algorithm.}
\label{fig:mergesort-2}
\end{figure}

The average time complexity of classic merge sort is $O(n \log(n))$, which is the same as quick sort and heap sort. Additionally, the best and worst case time complexity of merge sort is also $O(n \log(n))$, which is the also same as quicksort and heap sort. As a result, classical merge sort is generally unaffected by factors in the initial array.

However, classical merge sort uses $O(n)$ space, since additional memory is required when merging. Quicksort also has this space complexity, while heap sort takes $O(1)$ space, since it is an in-place method with no other memory requirements.

A summary of the complexities is shown in Table \ref{tab:complexity}.

\begin{table*}[htb]
\caption{Complexity of sort algorithms}
\label{tab:complexity}
\begin{tabular}{lllll}
    Sort & Average Time Complexity & Best Time Complexity & Worst Time Complexity \\
    \hline
    Bubble sort     & $O(n^2)$       & $O(n)$           & $O(n^2)$          \\
    Insertion sort  & $O(n^2)$       & $O(n)$           & $O(n^2)$          \\
    Selection sort  & $O(n^2)$       & $O(n^2)$         & $O(n^2)$          \\
    Heap sort       & $O(n\log(n))$  & $O(n\log(n))$    & $O(n\log(n))$    \\
    Quick sort      & $O(n\log(n))$  & $O(n\log(n))$    & $O(n\log(n))$   \\
    Merge sort      & $O(n\log(n))$  & $O(n\log(n))$    & $O(n\log(n))$     \\
    \end{tabular}
\end{table*}

\subsection{Python Built-in Sort} \label{builtin}

For the sake of benchmarking our merge sorts, we also run and compare the times of the Python built-in sort \lstinline{sorted()}. Python uses an algorithm called Timsort, a hybrid sorting algorithm of merge sort and insertion sort. The algorithm finds "runs", or subsequences of the data that are already order, and when these runs fulfill specific merge criteria, they are merged. 

The Python built-in sort is run by calling \lstinline{sorted(a)}, where a is the list to be sorted. 
\\
\subsection{Multiprocessing Merge Sort} \label{multiprocessing-ms}

Within a Symmetric Multiprocessing Machine (SMP), two or more identical processors are connected to a single mainstream memory, with full accessibility to all the input/output devices. In the case of multiple core machines, the SMP architecture treats each core as a separate processor.

To parallelize Python code on SMP machines, we use multiprocessing, which is a package that supports programming using multiple cores on a given machine. Python's Global Interpreter Lock (GIL) only allows one thread to have exclusive access to the interpreter resources, which means multithreading cannot be used when the Python interpreter is required. However, the multiprocessing package sidesteps this issue by spawning multiple processes instead of different threads. Because each process has its own interpreter with a separate GIL that carries out its given instructions, multiple processes can be run in parallel on one multicore machine. 

A common way to parallelize Python code is to employ a pool of processes. multiprocessing provides this with the \lstinline{Pool} class, which controls a pool of worker processes to which jobs can be submitted. There are four main steps in the usage of the \lstinline{Pool} class.

\begin{enumerate}
    \item Create the pool using
    \\
    \lstinline{pool = multiprocessing.Pool(processes=n)}

    Initializes a \lstinline{Pool} object \lstinline{pool} with \lstinline{n} worker processes. 
    \item Submit tasks.
    \begin{enumerate}
        \item Synchronous
        \item Asynchronous
    \end{enumerate}
    \item Wait for tasks to complete (dependent upon type of task submission). 
    \item Shut down the pool. \\
    pool.close()
\end{enumerate}

\subsubsection{Synchronous Tasks}

The \lstinline{Pool} provides a parallel version of the sequential \lstinline{map} function. Note that \lstinline{map} blocks, meaning that all worker processes must finish before the program can move on. 

\begin{figure}[htb]
\footnotesize
\begin{lstlisting}
  # example function
  def function(input):
    ...
    return result
    
  # input, divided into chunks
  input = [chunk1, chunk2, ...]
    
  # create pool with n processes
  pool = multiprocessing.Pool(processes=n)
    
  # start worker processes in parallel
  # output from each function call 
  # is appended to results
  results = pool.map(function, input)
\end{lstlisting}
\caption{Parallel function of map using Pool.}
\label{fig:para-map}
\end{figure}

\subsubsection{Asynchronous Tasks}

The \lstinline{Pool} also provides an asynchronous version of the sequential \lstinline{map} function called \lstinline{map_async()}. It does not block and returns a \lstinline{AsyncResult} to access later. 

\begin{figure}[htb]
\footnotesize
\begin{lstlisting}
  ...
  # start worker processes in parallel
  # issue tasks to the 
  # worker processes asynchronously
  results = pool.map_async(function, input)
  
  # iterate over the results from the tasks
  for value in results.get():
    ...
\end{lstlisting}
\caption{Parallel function of \lstinline{map_async()} using Pool.}
\label{fig:async}
\end{figure}

While parallelizing a program reduces the computational time by adding workers and reducing the size of the problem for each process, it also adds unavoidable overhead. Costs include the initialization time, overhead due to external libraries, and the cost of communication between workers. 

The official documentation for multiprocessing is located within the Python Standard Library \cite{multiproc}. 

A relatively straightforward conversion of sequential merge sort to multiprocessing merge sort can be done. Note that for this implementation, the array to be sorted and the number of processes to be used are passed in as parameters. 

To initialize the multiprocessing merge sort, we begin by initializing the Pool and dividing up the array.

\begin{figure}[htb]
\footnotesize
\begin{lstlisting}
  def multiprocessing_mergesort(arr, processes):
    # create processor pool 
    pool = multiprocessing.Pool(
           processes=processes)
        
    # split array into equally sized chunks
    # one for each process
    # append chunks to arr1
    size = int(math.ceil(
           float(len(arr)) / processes))
    arr1 = []
    for i in range(processes):
      arr1.append(arr[(size*i):(size*(i+1))])
\end{lstlisting}
\caption{Initialization of multiprocessing merge sort.}
\label{fig:init-mp-ms}
\end{figure}

Next, we spawn the worker processes, which run \lstinline{fast_sort} on each chunk in parallel. \lstinline{fast_sort} is the built-in Python \lstinline{sorted()} function. Note that we use synchronous communication here by using \lstinline{map()}.

\begin{figure}[htb]
\footnotesize
\begin{lstlisting}
  arr1 = pool.map(fast_sort, arr1)
  pool.close()
\end{lstlisting}
\caption{Parallel mapping from sort to array chunk.}
\label{fig:mapping}
\end{figure}

Finally, we merge the sorted chunks into one whole array.

\section{Analysis of Single Processor Sorting} \label{analysis1}

The performance of all sorts was measured on Carbonate, Indiana University's large-memory computer cluster \cite{carbo}. Sorts were ran on the general-purpose compute nodes, each of which is a Lenovo NeXtScale nx360 M5 server equipped with two 12-core Intel Xeon E5-2680 v3 CPUs. Each node has 256 GB of RAM and runs under Red Hat Enterprise 7.x. The sequential merge sort and built-in sort were run on a single core, while the multiprocessing merge sort was run on varying numbers of cores, from 1 to 24. All sorts were run on arrays of varying sizes of up to $10^7$. \\

\begin{figure}[htb]

\begin{tabular}{lrlrrl}
\toprule
 c &     size &   sort &    time &  speedup & efficiency \\
\midrule
 1 & $10^7$ &     mp &  7.724 & 1.000 &        1.000 \\
 4 & $10^7$ &     mp &  3.474 & 2.223 &   0.556 \\
 8 & $10^7$ &     mp &  3.164 & 2.441 &   0.305 \\
12 & $10^7$ &     mp &  2.487 & 3.106 &    0.259 \\
16 & $10^7$ &     mp &  2.820 & 2.739 &   0.171 \\
20 & $10^7$ &     mp &  2.858 & 2.703 &   0.135 \\
24 & $10^7$ &     mp &  2.830 & 2.730 &   0.114 \\
\midrule
 1 & $10^7$ &    seq & 85.611 & --- &   --- \\
\midrule
 1 & $10^7$ & sorted &  3.860 & --- &   --- \\
\bottomrule
\end{tabular}

\caption{Performance results on one node from \\ Carbonate partition (all times in seconds, mp = \\multiprocessing, seq = sequential merge sort).}
\label{fig:table-seq-1}
\end{figure}

Figure \ref{fig:table-seq-1} displays the performance results of the three single-machine sorting algorithms: multiprocessing merge sort, sequential merge sort, and the built-in Python \lstinline{sorted()}. The multiprocessing merge sort is much slower than the built-in sort at the beginning when running on one core. However, once two cores are reached, the multiprocessing merge sort becomes the fastest sort, and continues to get faster as the number of cores increases. The multiprocessing merge sort achieves a 34x speedup and the built-in Python sort achieves a 22x speedup in comparison to the sequential merge sort. The multiprocessing merge sort typically outperforms the built-in Python sort, achieving a 1.5x speedup when measured using 12 cores. 

\begin{figure}[h]
\includegraphics[width=1.0\columnwidth]{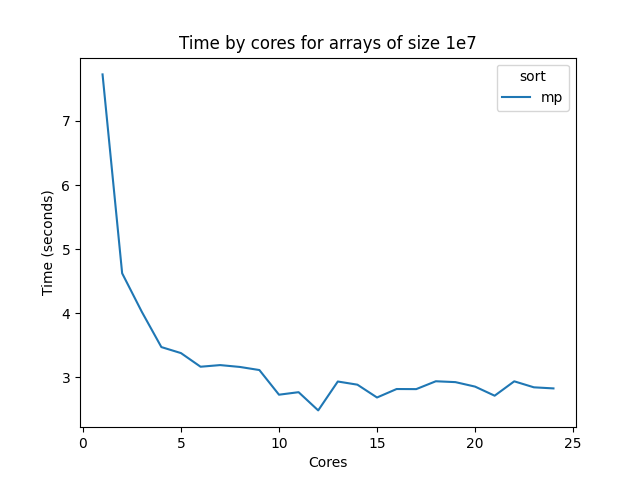}
\caption{Time by number of cores for arrays of size $10^7$.}
\label{fig:time-by-c-mp-1e7}
\end{figure}

For large arrays, increased parallelism is a negative factor for time. Figure \ref{fig:time-by-c-mp-1e7} displays the relationship between the number of cores used and the time of the multiprocessing merge sort algorithm for arrays of size $10^7$. 

\begin{figure}[ht]
\includegraphics[width=1.0\columnwidth]{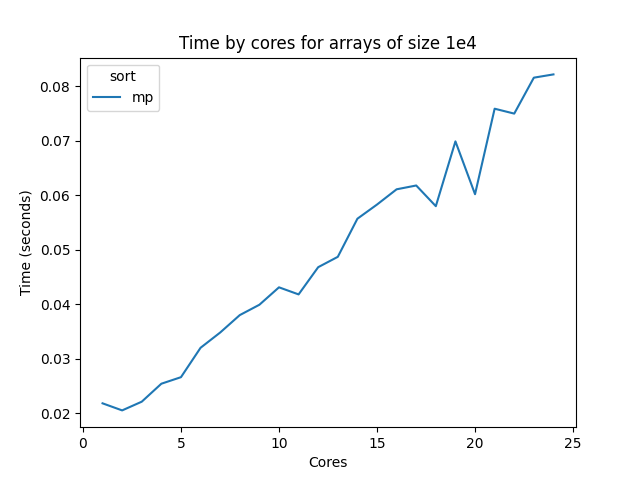}
\caption{Time by number of cores for arrays of size $10^4$.}
\label{fig:time-by-c-mp-1e4}
\end{figure}

However, this relationship does not hold true for smaller arrays. Figure \ref{fig:time-by-c-mp-1e4} displays the cores and times of the multiprocessing merge sort algorithm for arrays of size $10^4$. In this case, increased parallelism is actually a positive factor for time. Since the array is small, the ratio of the code that is parallelized to the code that must run sequentially is also small. This means that the reductions in speed parallelization are not large enough to offset the parallelization overhead. While costs like initialization and overhead from external libraries remain constant, the cost of communication between processes increases significantly with the number of processes. 

\begin{figure}[ht]
\includegraphics[width=1.0\columnwidth]{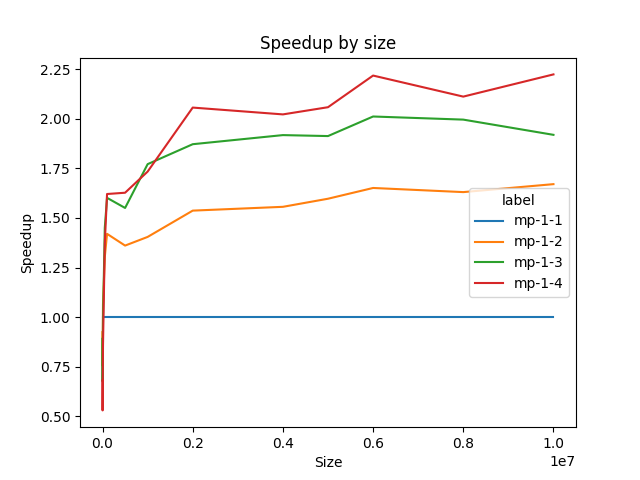}
\caption{Speedup by size for 1, 2, 3, and 4 cores (mp-1-$c$ means multiprocessing merge sort on $c$ cores.)}
\label{fig:speedup-by-size-self-1234}
\end{figure}

In terms of speedup, gains are made extremely quickly when (a) the number of cores being used is small, and (b) the size of the array is small. As the experiments increase from using 1 core to using 4 cores, we achieve substantial increases in speedup, as can be seen in Figure \ref{fig:speedup-by-size-self-1234}. Furthermore, speedup, which initially is less than one for small arrays, increases as the array sizes increase from 100 to around $2 \times 10^6$. However, we can see that as the array sizes continue to increase past that point, speedup begins to level off, increasing at a much slower rate. 

\begin{figure}[htb]
\includegraphics[width=1.0\columnwidth]{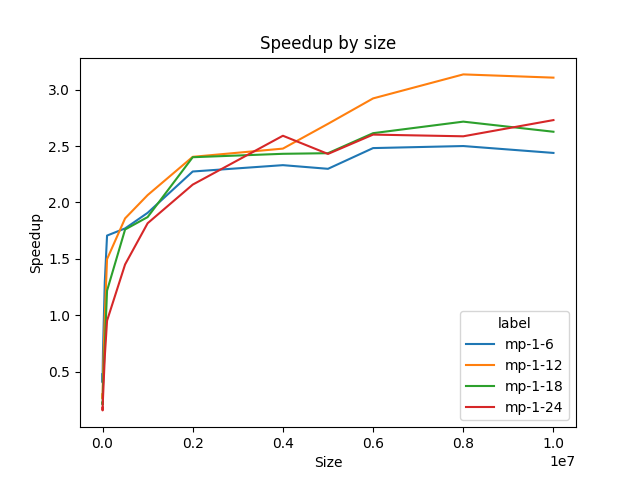}
\caption{Speedup by size for 6, 12, 18, and 24 cores (mp-1-$c$ means multiprocessing merge sort on $c$ cores.)}
\label{fig:speedup-by-size-self-1216}
\end{figure}

For larger numbers of cores, speedup also begins to level off, and differences in performance between the numbers of cores used become negligible. In Figure \ref{fig:speedup-by-size-self-1216}, multiprocessing merge sort on 6, 18, and 24 cores all display very similar speedup, despite the large difference in the number of cores.

Multiprocessing merge sort on 12 cores outperforms them all, but this can be attributed to system architecture. Since each node on the Carbonate has two 12-core Intel Xeon CPUs, 12 is the optimal number of cores on which to run multiprocessing merge sort. Any more cores requires cross-CPU communication between the two CPUs, a much higher cost than the intra-CPU communication done by 12 or less cores.  

\section{MPI Merge Sort} \label{mpi-ms}

The limitation of multiprocessing is that it does not support parallelism over distributed computing systems. To enable the communication of computing nodes, we use MPI for Python, also known as mpi4py. An object oriented approach to message passing in Python, it closely follows the MPI-2 C++ bindings. Within a distributed computing system, each node has its own local memory, and information can only be exchanged by passing messages through available communication links. The mpi4py module provides both point-to-point and collective communication capabilities for both Python objects and buffer-like objects. 

Some usages of mpi4py are introduced:
By default, there is a single group that contains all CPUs. \lstinline{MPI.COMM_WORLD} is the intra-communicator that handles cross-node communication. \lstinline{comm.Get_size()} is used to obtain the number of CPUs available in the group to perform the computation, and \lstinline{comm.Get_rank()} is used to obtain the rank of the current CPU. The rank of a CPU is its unique numerical identifier. CPUs can have rank 0 through (size - 1), the assignment of which is entirely random. Typically, the root node will be designated as whichever CPU has rank 0, and ranks 1 through (size - 1) will be the secondary nodes. 

Point-to-point communication is carried out using \lstinline{comm.send()} and \lstinline{comm.recv()}. As described in the name, comm.send is used to send data from one CPU, and comm.recv is used to receive data on the recipient CPU. Each envelope contains:

\begin{enumerate}
    \item The data that is to be sent
    \item The datatype of each element of the data
    \item The size of the data
    \item An identification number for the message
    \item The ranks of the sending and receiving nodes
\end{enumerate}

Senders and receivers in a program should be matched, and there should be one receive per send. The sending process exits once the send buffer can be read and written to, and the receiving process exits once it has successfully received the message in the receive buffer. Since completion of communication depends on both processes in the pair, there is a risk that if one fails, the program will be stuck indefinitely.

Collective communication transmits data to every CPU in the group. Important functions for collective communication in mpi4py are:
\begin{enumerate}
    \item Broadcasting:\\
    \lstinline{comm.Bcast(data, root=0)}
    
    The root node sends data to all other CPUs in the group.
    \item Scattering: \\
    \lstinline{comm.Scatter(sendbuf, recvbuf, root=0)}
    
    The root node sends equal chunks of the data (sendbuf) to all other CPUs in the group. 
    \item Gathering: \\
    \lstinline{comm.Gather(sendbuf, recvbuf, root=0)}
    
    All of the data from all CPUs in the group is collected to the root node. 
\end{enumerate}

Note that with mpi4py, the communication of generic Python objects utilizes all-lowercase names like comm.send(), comm.recv(), and comm.bcast(), while the communication of buffer-like objects utilizes upper-case names like comm.Send(), comm.Recv(), and comm.Bcast().

All MPI processes begin at the same time once spawned, and all programs execute the same code in parallel. Therefore, each process must be able to recognize its own position and role within the process tree. MPI processes must use their ranks to map themselves to nodes, forming a virtual tree. The process with rank 0 (hereinafter referred to as process 0 for brevity) is the root of the tree, with the other processes being the nodes of the tree. 

The official documentation is hosted by Read the Docs \cite{mpi}. 

All MPI processes run the same code, which must identify  the root process and the helper processes. The root process (process 0) generates the unsorted array that is to be sorted and then sends chunks of the unsorted array to each helper process using comm.Scatter(). Each helper process (i) receives its local array from the root process, (ii) invokes the sorting algorithm determined in part 1; and (iii) merges its local array with a paired node. 

\subsection{Dividing array into subarrays.}

We generate the unsorted array as a NumPy array on process 0. The array is scattered to all the processes so each process has an equal-sized chunk of the list (or subarray). Note that this means the size of the array must be evenly divisible by the number of processes. 

\begin{figure}[htb]
\includegraphics[width=1.0\columnwidth]{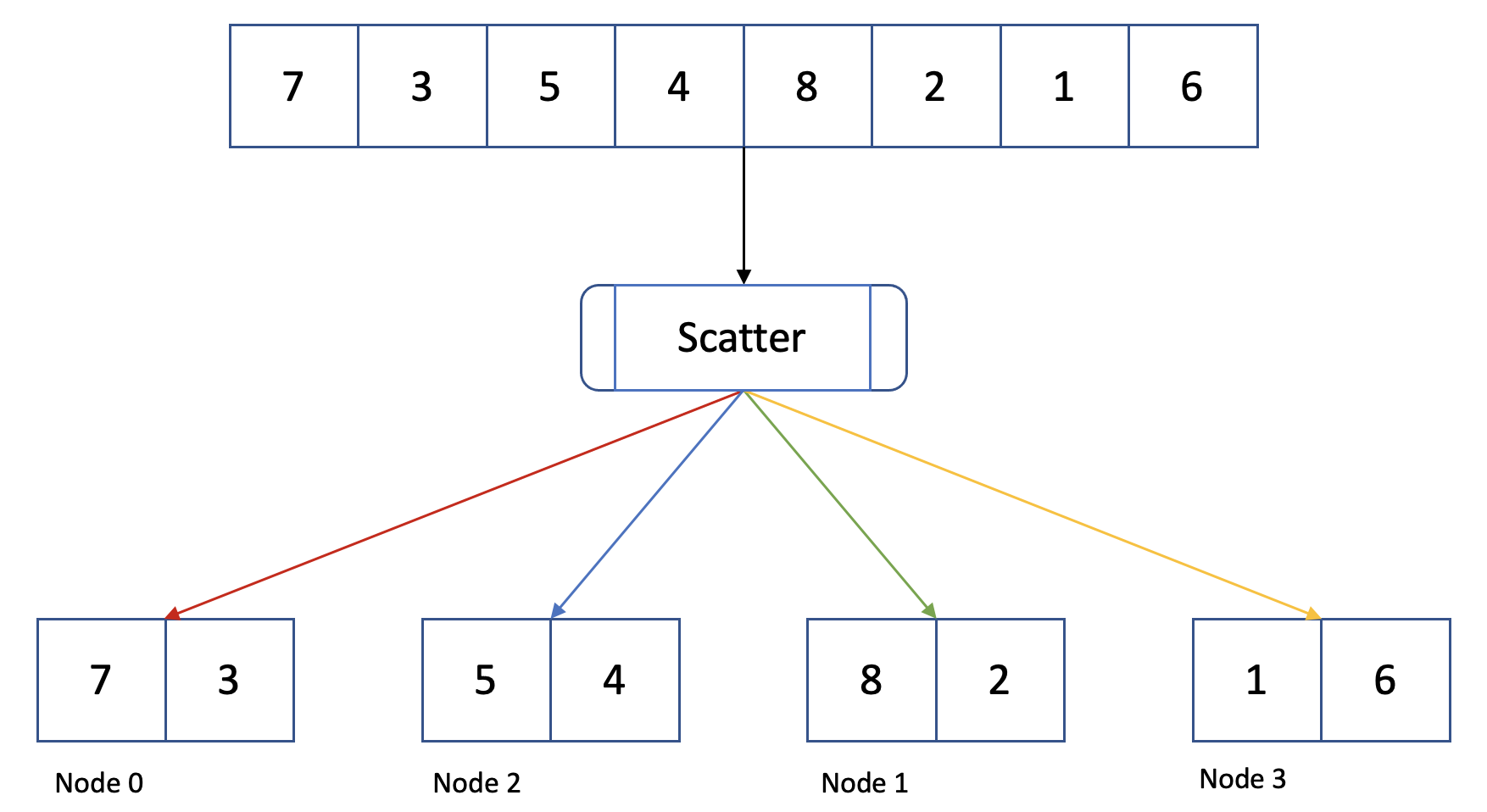}
\caption{Example of MPI Scatter.}
\label{fig:mpi-scatter}
\end{figure}

Figures \ref{fig:mpi-scatter} and \ref{fig:mpi-scatter-code} assume that we have 4 processes and an array of size 8. 

\begin{figure}[htb]
\footnotesize
\begin{lstlisting}
  # set up MPI
  comm = MPI.COMM_WORLD
  size = comm.Get_size()
  rank = comm.Get_rank()
    
  # size of each subarray
  sub_size = int(n / size)
    
  # allocate local memory on each rank
  local = np.zeros(sub_size, dtype="int")
     
  if rank is 0:
    # generate unsorted array
    unsorted = np.random.randint(n, size=n)
    
  comm.Scatter(unsorted, local, root=0)
\end{lstlisting}
\caption{Python code implementation of MPI Scatter.}
\label{fig:mpi-scatter-code}
\end{figure}

\subsection{Sorting subarrays.}
Once the subarrays have been distributed using the Scatter command, they are sorted on each processor. The specific sorting algorithm used to do the intra-node sorting will be known as the subsort. The implementation of two different types of MPI merge sorts is carried out, and the type of MPI merge sort is dependent upon the subsort algorithm. The options for the subsort algorithm are:

\begin{enumerate}
    \item Built-in Python sort, where the subarray is sorted using \lstinline{sorted()} (See Section \ref{builtin}). We will refer to this as message-passing merge sort. 
    \item Multiprocessing merge sort, where the subarray is sorted using multiprocessing\_mergesort (See Section \ref{multiprocessing-ms}). We will refer to this as hybrid MPI merge sort. 
\end{enumerate}

\begin{figure}[htb]
\includegraphics[width=1.0\columnwidth]{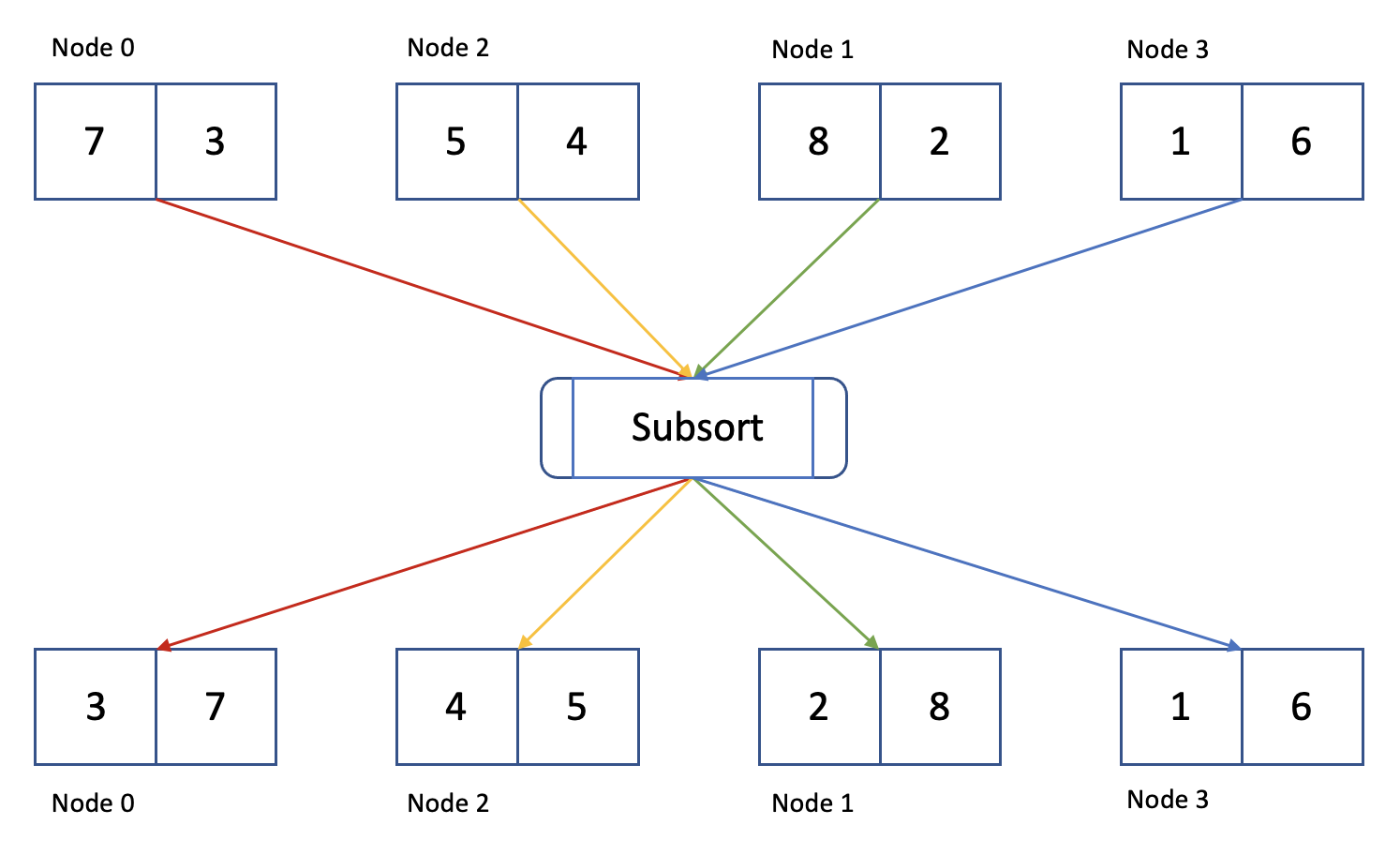}
\caption{Example of sorting subarrays using subsort.}
\label{fig:mpi-sort-ex}
\end{figure}

\subsection{Merging the subarrays.}

To merge the sorted subarrays, we enter a loop. Each loop, we will remove the bottom-most layer of the process tree, e.g. the leaves of the tree. The loop will continue until we have a single sorted list. 

\begin{figure}[htb]
\includegraphics[width=1.0\columnwidth]{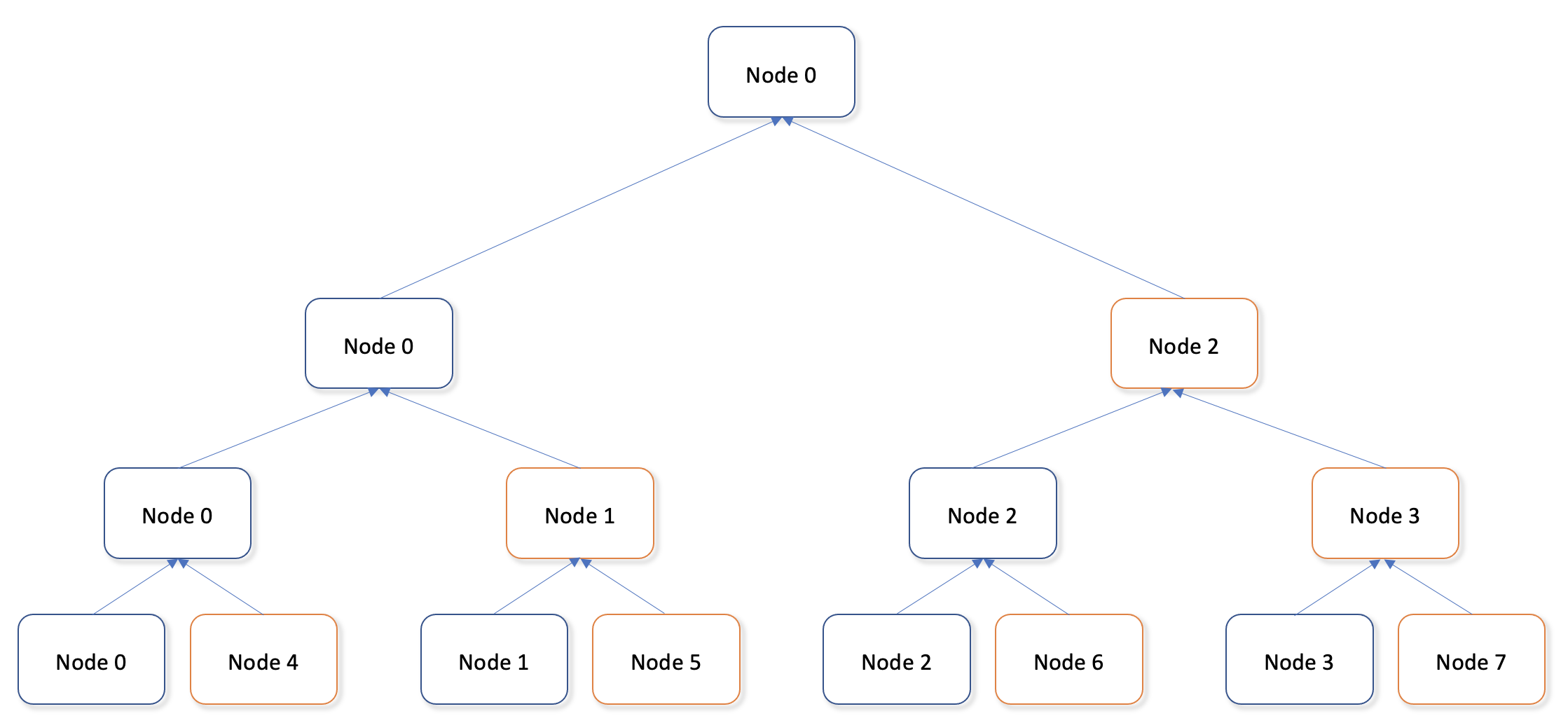}
\caption{Example process tree.}
\label{fig:rank-tree}
\end{figure}

Within the loop, we determine whether the current process is a left child or a right child of the parent. Note that a left child and its parent node will be the same process. In Figure \ref{fig:rank-tree}, the right child nodes are outlined in orange. For example, we can see that process 0 is the left child and process 1 is the right child of parent process 0. 

The way that parent and child nodes are determined in the tree is by process rank. The lower half ranks are the parent processes (and left children) and the upper half ranks are the right children. For example, in the figure above, we can see that ranks 0, 1, 2, and 3 are the parent/left child processes, and ranks 4, 5, 6, and 7 are the right child processes. 

Depending on the rank of the process, the node will classify itself as a left child or right child, and it will do the following:
\\

Left child :
\begin{enumerate}
    \item Allocate memory needed for storing right child's array in tmp
    \item Allocate memory needed for result of merging lists in result
    \item Receive right child's array from process $rank + split$, where $split$ is the current number of processes divided by two. For example, process 1 knows to receive data from process $1 + 4$, or process 5. 
    \item Merge local array with tmp in result\\
\end{enumerate}

Right child:
\begin{enumerate}
    \item Send local array to process $rank - split$, where $split$ is the current number of processes divided by two. For example, process 5 knows to send its data to process $5 - 4$, or process 1. 
\end{enumerate}

\begin{figure}[htb]
\includegraphics[width=1.0\columnwidth]{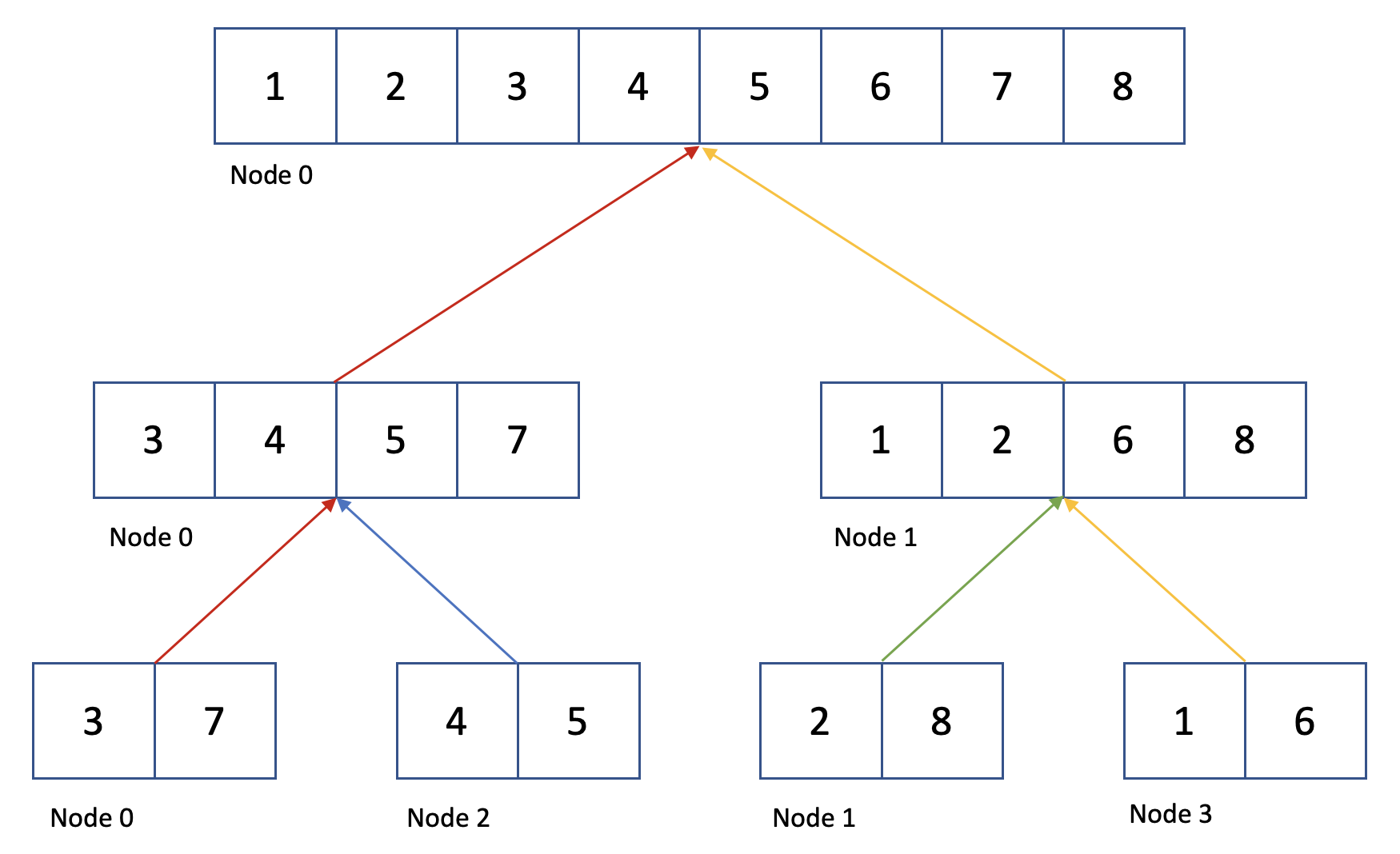}
\caption{Example of MPI merging.}
\label{fig:mpi-merge}
\end{figure}

By dividing nodes in half and mapping between the halves, unique pairings of parent and child processes are guaranteed for every step of the loop. Figure \ref{fig:mpi-merge} provides an example of merging sorted subarrays in each step of the process tree. 

Figure \ref{fig:mpi-merge-code} shows the Python implementation.

\begin{figure}[htb]
\footnotesize
\begin{lstlisting}
  split = size / 2
  while split >= 1:
    # rank is in upper half of processes
    # this process is a right child
    if split <= rank < split * 2:
      # send local array 
      # to parent to be merged
      comm.Send(local, rank-split, tag=0)
        
    # rank is in lower half of processes
    # this process is a left child/parent
    elif rank < split:
      # allocate memory 
      # for right child's array
      tmp = np.zeros(local.size, dtype="int")
      # allocate memory for merged result
      result = 
      np.zeros(2*local.size, dtype="int")
            
      # receive data from right child
      comm.Recv(tmp, rank+split, tag=0)
            
      # merge arrays
      result = merge(local, tmp)
      local = np.array(result)
    # update split as we have removed 
    # bottom layer of process tree
    # or half of the nodes
    split = split / 2
\end{lstlisting}
\caption{Python code example of MPI merging.}
\label{fig:mpi-merge-code}
\end{figure}

\section{Analysis of MPI Sorting} \label{analysis2}

The performance of all sorts was measured on Carbonate, Indiana University's large-memory computer cluster. Specific hardware details can be viewed in section \ref{analysis1}. Message-passing merge sort (with MPI, Section \ref{mpi-ms}) was executed on 1, 2, and 4 general processing nodes by using one single core on all nodes with MPI processes. Hybrid MPI merge sort was executed on 1, 2, and 4 nodes by using all available cores on all nodes for distributed MPI processes. All sorts were run on arrays of varying sizes of up to $10^7$. 

To see a detailed breakdown of the performances of the merge sorts using MPI, please see Figure \ref{fig:three-graphs}. Overall, message-passing merge sort runs significantly faster than the hybrid merge sort, and runs slightly slower than multiprocessing merge sort does. Notably, the MPI merge sort using multiprocessing merge sort performs significantly worse than the MPI merge sort using sorted, despite the multiprocessing merge sort outperforming sorted in Section \ref{analysis1}. This can be attributed to the usage of fork() system calls by the multiprocessing library to create worker processes, which is not compatible with some MPI implementations, including mpi4py. However, sorted runs completely sequentially and never faces this obstacle, resulting in the difference in time. 
Increased parallelism between nodes significantly speeds up run time, as is to be expected. However, due to the limited number of nodes available, there is not enough data to project the potential sort performance with further increased parallelism. 

\begin{figure} [htb]
\begin{tabular}{rrrllllr}
\toprule
 p &  c &     size & sort & sub & user & node &    time \\
\midrule
 1 & 24 & $10^7$ &  mpi &      mp & alex & v100 & 71.227 \\
 2 & 24 & $10^7$ &  mpi &      mp & alex & v100 & 37.543 \\
 4 & 24 & $10^7$ &  mpi &      mp & alex & v100 & 21.622 \\
\bottomrule
\end{tabular}
\caption{Hybrid MPI merge sort with varying processes (sub = subsort).}
\label{fig:mpi-time-p-24}
\end{figure}

\begin{figure}[htb]
    \centering
    \includegraphics[width=1.0\columnwidth]{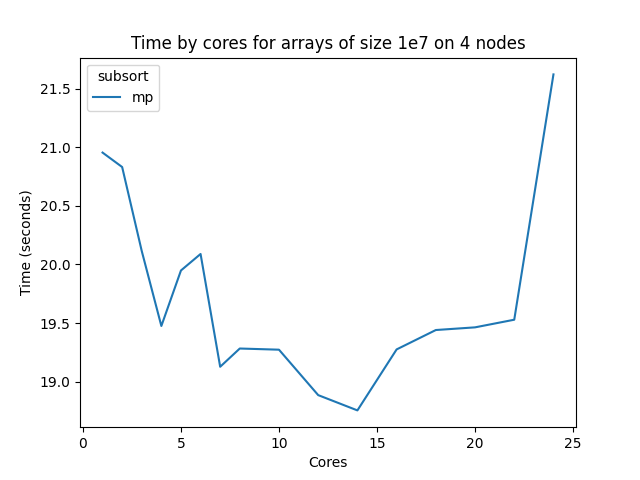}
    \caption{Time by cores for hybrid MPI merge sort on 4 nodes and array size $10^7$.}
    \label{fig:mpi-time-c-4-1e7}
\end{figure}

Increased parallelism within each node does appear to correlate with faster run times, but only up to a certain point. For example, in Figure \ref{fig:mpi-time-c-4-1e7}, decreases in time are seen up until 14 cores are used, at which point parallelism becomes correlated with increasing run times. 

\begin{figure}[htb]
    \centering
    \includegraphics[width=1.0\columnwidth]{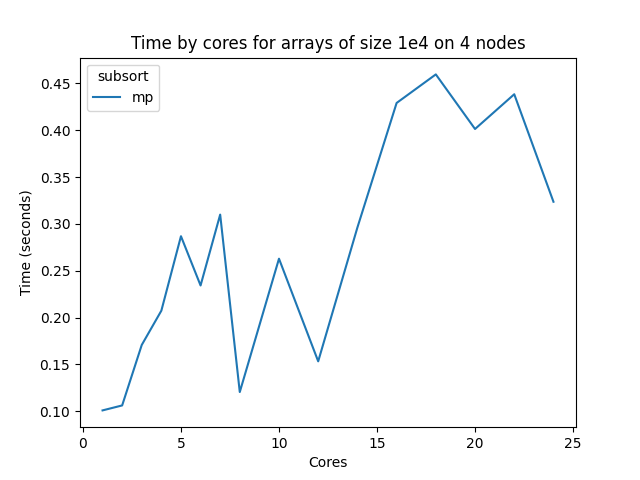}
    \caption{Time by cores for hybrid MPI merge sort on 4 nodes and array size $10^4$.}
    \label{fig:mpi-time-c-4-1e4}
\end{figure}

For arrays of smaller sizes, the relationship between intra-node parallelism and speed is much more variable and cannot be quantified. Figure \ref{fig:mpi-time-c-4-1e4} shows the extreme variation in time for arrays of size $10^4$. However, the positive relationship between parallelism and run time after a certain point (around 12 cores for this example) still holds. 

Finally, in comparison to the multiprocessing merge sort, the message-passing merge sort underperforms. As mpi4py runs on a distributed memory system, it must make a copy of every message that it sends, resulting in a large overhead for every message that must be sent. On the other hand, since the multiprocessing merge sort runs on a SMP, message passing between cores is much faster due to the lighter memory requirements. 

\begin{figure}[htb]
    \centering
    \includegraphics[width=1.0\columnwidth]{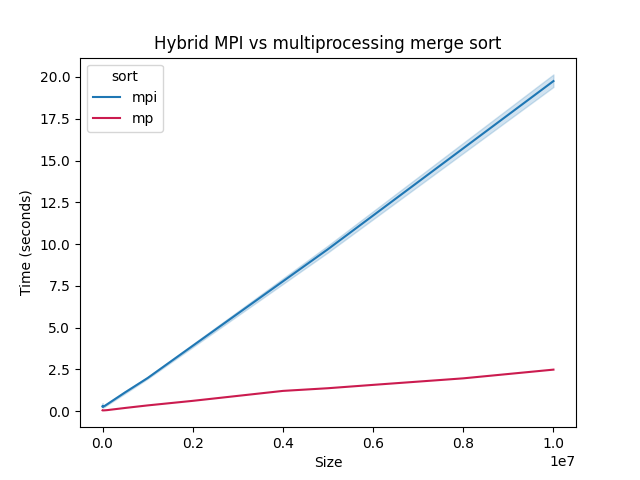}
    \caption{Time by size compared between message-passing merge sort on 4 nodes and multiprocessing merge sort.}
    \label{fig:hybrid-vs-mpi}
\end{figure}

\section{Conclusion} \label{end}

This paper introduces and implements three parallel merge sort algorithms in the Python programming language: multiprocessing merge sort, message-passing merge sort, and hybrid MPI merge sort. We run each sorting algorithm on the Carbonate computing cluster, and using Cloudmesh, we benchmark, compare, and analyze the results of the sorting algorithms. We also measure the performance sequential merge sort and the built-in Python sort to act as a benchmark for the parallel sorts. 

We find that while running on our machine, for arrays with sizes within the range of $2 \times 10^6$ and $10^7$, multiprocessing merge sort is the fastest sorting algorithm, followed by the message-passing merge sort, and then the hybrid MPI merge sort. We discover that multiprocessing merge sort outperforms even the built-in Python sort, running 1.5x faster in some cases. However, for small array sizes, the multiprocessing merge sort is relatively slow due to the associated overhead. We conclude that while for smaller problems, it may be more beneficial to use a sequential solution, parallelism should be implemented once the problem size is sufficiently large. Note that the best parallel algorithm to use, along with the threshold for {\em sufficiently large}, depends upon the nature of the problem, the network type, and the hardware and software present.

\section{Acknowledgements}

The author of the paper is thankful to Dr. Gregor von Laszewski, University of Virginia for the introduction in benchmarking \cite{las-stopwatch}, parallel computing concepts, and MPI for Python \cite{cloudmesh-mpi}.

\bibliographystyle{IEEEtran}
\bibliography{main}

\begin{thebibliography}{10}
\providecommand{\url}[1]{#1}
\csname url@samestyle\endcsname
\providecommand{\newblock}{\relax}
\providecommand{\bibinfo}[2]{#2}
\providecommand{\BIBentrySTDinterwordspacing}{\spaceskip=0pt\relax}
\providecommand{\BIBentryALTinterwordstretchfactor}{4}
\providecommand{\BIBentryALTinterwordspacing}{\spaceskip=\fontdimen2\font plus
\BIBentryALTinterwordstretchfactor\fontdimen3\font minus
  \fontdimen4\font\relax}
\providecommand{\BIBforeignlanguage}[2]{{%
\expandafter\ifx\csname l@#1\endcsname\relax
\typeout{** WARNING: IEEEtran.bst: No hyphenation pattern has been}%
\typeout{** loaded for the language `#1'. Using the pattern for}%
\typeout{** the default language instead.}%
\else
\language=\csname l@#1\endcsname
\fi
#2}}
\providecommand{\BIBdecl}{\relax}
\BIBdecl

\bibitem{selkie}
\BIBentryALTinterwordspacing
``{CS in Parallel}.'' [Online]. Available:
  \url{http://selkie.macalester.edu/csinparallel/modules/ParallelSorting/build/html/MergeSort/MergeSort.html}
\BIBentrySTDinterwordspacing

\bibitem{cole88}
\BIBentryALTinterwordspacing
R.~Cole, ``Parallel merge sort,'' \emph{SIAM Journal on Computing}, vol.~17,
  no.~4, pp. 770--785, 1988. [Online]. Available:
  \url{https://doi.org/10.1137/0217049}
\BIBentrySTDinterwordspacing

\bibitem{jeon03}
M.~Jeon and D.~Kim, ``{Parallel Merge Sort with Load Balancing},''
  \emph{International Journal of Parallel Programming}, vol.~31, pp. 21--33, 02
  2003.

\bibitem{mars18}
Z.~Marszałek, M.~Woźniak, and D.~Polap, ``{Fully Flexible Parallel Merge Sort
  for Multicore Architectures},'' \emph{Complexity}, vol. 2018, 12 2018.

\bibitem{uyar14}
A.~Uyar, ``{Parallel merge sort with double merging},'' \emph{2014 IEEE 8th
  International Conference on Application of Information and Communication
  Technologies (AICT)}, 10 2014.

\bibitem{odeh}
S.~Odeh, O.~Green, Z.~Mwassi, O.~Shmueli, and Y.~Birk, ``Merge path - parallel
  merging made simple,'' in \emph{2012 IEEE 26th International Parallel and
  Distributed Processing Symposium Workshops}, 2012, pp. 1611--1618.

\bibitem{davidson}
A.~Davidson, D.~Tarjan, M.~Garland, and J.~D. Owens, ``Efficient parallel merge
  sort for fixed and variable length keys,'' in \emph{2012 Innovative Parallel
  Computing (InPar)}, 2012, pp. 1--9.

\bibitem{radenski11}
\BIBentryALTinterwordspacing
A.~Radenski, ``{Shared Memory, Message Passing, and Hybrid Merge Sorts for
  Standalone and Clustered SMPs},'' \emph{2011 International Conference on
  Parallel and Distributed Processing Techniques and Applications}, pp.
  367--373, 01 2011. [Online]. Available:
  \url{https://digitalcommons.chapman.edu/scs_books/19/}
\BIBentrySTDinterwordspacing

\bibitem{cloudmesh-mpi}
\BIBentryALTinterwordspacing
``{Cloudmesh with MPI},'' 2022. [Online]. Available:
  \url{https://github.com/cloudmesh/cloudmesh-mpi}
\BIBentrySTDinterwordspacing

\bibitem{las-stopwatch}
\BIBentryALTinterwordspacing
G.~von Laszewski, J.~P. Fleischer, and G.~C. Fox, ``Hybrid reusable
  computational analytics workflow management with cloudmesh,'' 2022. [Online].
  Available: \url{https://arxiv.org/abs/2210.16941}
\BIBentrySTDinterwordspacing

\bibitem{multiproc}
\BIBentryALTinterwordspacing
``{multiprocessing - Process-based parallelism},'' 2022. [Online]. Available:
  \url{https://docs.python.org/3/library/multiprocessing.html}
\BIBentrySTDinterwordspacing

\bibitem{carbo}
\BIBentryALTinterwordspacing
``{About Carbonate at Indiana University},'' 2022. [Online]. Available:
  \url{https://kb.iu.edu/d/aolp}
\BIBentrySTDinterwordspacing

\bibitem{mpi}
\BIBentryALTinterwordspacing
L.~Dalcin, ``{MPI for Python},'' 2022. [Online]. Available:
  \url{https://mpi4py.readthedocs.io/en/stable/}
\BIBentrySTDinterwordspacing

\end{thebibliography}

\clearpage

\section{Appendix}

\begin{figure}[htb]
\centering
    \begin{subfigure}[b]{\linewidth}
    \centering
    \begin{tabular}{rrrllr}
    \toprule
     p &  c &     size & sort & subsort &    time \\
    \midrule
     1 &  1 & $10^7$ &  mpi &  sorted &  4.237 \\
    \midrule
     1 &  1 & $10^7$ &  mpi &      mp & 80.335 \\
     1 &  4 & $10^7$ &  mpi &      mp & 78.716 \\
     1 &  8 & $10^7$ &  mpi &      mp & 77.275 \\
     1 & 12 & $10^7$ &  mpi &      mp & 76.570 \\
     1 & 16 & $10^7$ &  mpi &      mp & 77.588 \\
     1 & 20 & $10^7$ &  mpi &      mp & 78.152 \\
     1 & 24 & $10^7$ &  mpi &      mp & 71.227 \\
    \bottomrule
    \end{tabular}
    \caption{Performance results on 1 node.}
    \label{fig:table-mpi-1}
    \end{subfigure}
    
    \begin{subfigure}[b]{\linewidth}
    \centering
    \begin{tabular}{rrrllr}
    \toprule
     p &  c &     size & sort & subsort &    time \\
    \midrule
     2 &  1 & $10^7$ &  mpi &      mp & 38.846 \\
     2 &  4 & $10^7$ &  mpi &      mp & 38.508 \\
     2 &  8 & $10^7$ &  mpi &      mp & 36.842 \\
     2 & 12 & $10^7$ &  mpi &      mp & 37.975 \\
     2 & 16 & $10^7$ &  mpi &      mp & 37.800 \\
     2 & 20 & $10^7$ &  mpi &      mp & 39.400 \\
     2 & 24 & $10^7$ &  mpi &      mp & 37.543 \\
    \bottomrule
    \end{tabular}
    \caption{Performance results on 2 nodes.}
    \label{fig:table-mpi-2}
    \end{subfigure}
    
    \begin{subfigure}[b]{\linewidth}
    \centering
    \begin{tabular}{rrrllr}
    \toprule
     p &  c &     size & sort & subsort &      time \\
    \midrule
     4 &  1 & $10^7$ &  mpi &      mp & 20.954 \\
     4 &  4 & $10^7$ &  mpi &      mp & 19.475 \\
     4 &  8 & $10^7$ &  mpi &      mp & 19.282 \\
     4 & 12 & $10^7$ &  mpi &      mp & 18.885 \\
     4 & 16 & $10^7$ &  mpi &      mp & 19.275 \\
     4 & 20 & $10^7$ &  mpi &      mp & 19.463 \\
     4 & 24 & $10^7$ &  mpi &      mp & 21.622 \\
    \bottomrule
    \end{tabular}
    \caption{Performance results on 4 nodes.}
    \label{fig:table-mpi-4}
    \end{subfigure}
    \caption{Results from MPI merge sorts on various nodes with array size of $10^7$. }
    \label{fig:three-graphs}
\end{figure}

\end{document}